\RequirePackage{fix-cm}
\documentclass[twocolumn,epjc3]{svjour3}  
\smartqed  
\RequirePackage{graphicx}
\journalname{Eur. Phys. J. C}
\usepackage{soul}
\usepackage{graphicx}
\usepackage{enumerate}
\usepackage{pdfpages}
\usepackage{compactbib}
\usepackage{amssymb}
\usepackage{comment}
\usepackage{siunitx}
\usepackage{isotope}
\usepackage{url}
\usepackage{wrapfig}
\usepackage{cite}
\usepackage[outercaption]{sidecap}    
\usepackage{xspace} 
\usepackage[usenames,dvipsnames]{}
\usepackage[T1]{fontenc}
\usepackage[utf8]{inputenc}
\usepackage{longtable} %Mandatory for the tab
\usepackage{booktabs} %Mandatory for the tab
\usepackage{threeparttable} %Needed for footnotes in table
\usepackage{array} %Mandatory for the tab
\usepackage{caption}
\usepackage{subcaption}
\usepackage{xcolor}
\usepackage{epstopdf}
\usepackage{amsmath}
\usepackage[integrals]{wasysym}
\usepackage{multirow}
\usepackage[title]{appendix}%

\begin{document}

\title{The Gator low-background spectrometer: readout upgrade and photomultiplier tube radioassay for XLZD}
\titlerunning{Gator readout upgrade}        

\author{Laura Baudis\thanksref{e1}
        \and
        Sofia Calgaro
         \and
        Jose Javier Cuenca-Garc\'ia
        \and
       Florian J\"org
        \and
        Margherita Noia\thanksref{e2}
        \and
        Francesco Piastra
        \and
        Aravind Remesan Sreekala\thanksref{e3}
}
\thankstext{e1}{e-mail: laura.baudis@uzh.ch}
\thankstext{e2}{e-mail: margherita.noia@uzh.ch}
\thankstext{e3}{e-mail: aravind.sreekala@uzh.ch}
%\thankstext{e0}{e-mail: sofia.calgaro@uzh.ch}
%\thankstext{e0}{e-mail: jose.cuenca@uzh.ch}
%\thankstext{e0}{e-mail: florian.joerg@uzh.ch}
%\thankstext{e0}{e-mail: francesco.piastra@uzh.ch}
\institute{Physik-Institut, University of Zurich, Winterthurerstrasse 190, 8057 Zurich, Switzerland %\label{addr1}
}
\date{Received: date / Accepted: date}

\maketitle

\begin{abstract}
We report on a readout upgrade of Gator, a low-background high-purity germanium spectrometer operated at the Laboratori Nazionali del Gran Sasso for material radioassay. The upgraded data acquisition records digitised waveforms, enabling noise rejection on an event-level basis, and offline energy reconstruction. Measurements before and after the upgrade show an improved energy resolution of 17\% at the 2614.5\,keV line of $^{208}\text{Tl}$, with $(2.14 \pm 0.22)$\,keV at FWHM, and suppression of electronic noise in the low-energy region. We assess the resulting sensitivity for a specified sample geometry, and present a new background measurement. The background rate in the (100-2700)\,keV range is $(86.6 \pm 0.5$)\,counts/(kg\,day). We also show the radioassay results of sixteen Hamamatsu R17317 3-inch photomultiplier tubes with an improved stem design, which reduces the activities from the primordial $^{238}$U and $^{232}$Th chains by a factor of three, making them suitable candidate photosensors for the XLZD experiment. 

\end{abstract}

\section{Introduction}
\label{sec:intro}

Traces of radioactive isotopes in detector materials contribute to the backgrounds of dark matter, neutrinoless double beta decay and other rare--event searches, and 
material radioassay forms part of the selection of components for these type of experiments~\cite{Heusser:1995wd}.
High-purity germanium (HPGe) spectrometry offers a non--destructive measurement
of $\gamma$-emitting radionuclides, including daughters of the $^{238}$U and
$^{232}$Th decay chains, $^{40}$K, as well as cosmogenic and anthropogenic
isotopes. The sensitivity of a measurement depends
on the detector's background level and energy resolution, the counting time, and the
detection efficiency for the particular sample geometry and composition.

The Gator facility at the Laboratori Nazionali del Gran Sasso (LNGS) of INFN,
Italy, operates a 2.2\,kg, p-type coaxial HPGe detector at an average depth of
3600\,m water equivalent~\cite{Gator2011,Gator2022}. Passive shielding combined with a nitrogen purged enclosure, suppress external radiation and radon-induced backgrounds. Gator has been used for material screening for the XENON programme~\cite{Aprile:2011ru,XENON:2017fdb,XENON:2021mrg}, 
for GERDA~\cite{GERDA:2012qwd} and LEGEND-200~\cite{LEGEND:2025jwu}.
In addition, its low background has enabled a dedicated search for Pauli exclusion
principle violations in lead~\cite{Gator2024}. The detector construction,
calibration and efficiency validation were described in Ref.~\cite{Gator2011}.
The subsequent enclosure and sample-handling upgrade, together with updated
performance and radioassay methods, was reported in Ref.~\cite{Gator2022}.

In this work we detail an upgrade of the readout electronics and data
acquisition. The previous acquisition recorded energy spectra without
retaining individual waveforms. The addition of waveform digitisation provides
access to pulse and baseline information for investigating electronic noise,
applying data quality selections and reconstructing the deposited energy
at the software level. We characterise the detector after this upgrade through measurements of its energy resolution and background spectrum, and evaluate the sensitivity for a specified sample geometry. We also present the radioassay of sixteen Hamamatsu R17317 3-inch photomultiplier tubes with low-activity stems,
investigated as candidate photosensors for the XLZD experiment~\cite{BaudisXLZD2024,XLZDDesignBook}. Their activities are estimated
using a joint Poisson likelihood fit for sample and background measurements and
compared with results from the previously applied counting analysis.

In Section~\ref{sec:hardware} we describe the facility and readout upgrade.
The measurements and analysis methods are presented in
Section~\ref{sec:methods}, followed by the detector performance in
Section~\ref{sec:performance}. In section~\ref{sec:samples} we report the  3-inch PMT radioassay and compare the results with previous measurements of an earlier model. Section~\ref{sec:conclusions} concludes the manuscript and provides an outlook.

\section{The Gator facility and readout upgrade}
\label{sec:hardware}

In this section  we describe the facility and the upgrade of the data acquisition system. We motivate the hardware changes and describe waveform acquisition and energy reconstruction algorithm.

\subsection{Detector and shielding}
\label{sec:detector}

The HPGe detector is housed in an oxygen-free copper cryostat and cooled by a U-shaped
copper cold finger immersed in liquid nitrogen. Its relative full-energy peak
efficiency is 100.5\%\footnote{The efficiency is defined at the 1.33\,MeV $^{60}$Co line with respect to a
3-inch $\times$ 3-inch NaI(Tl) detector at a source distance of
25\,cm}. While this reference efficiency characterises the
detector, the efficiencies used for sample radioassay are calculated for each
measurement geometry, as we discuss in Sect.~\ref{sec:efficiency}.

\begin{figure}[tbp]
\centering
\includegraphics[width=\columnwidth]{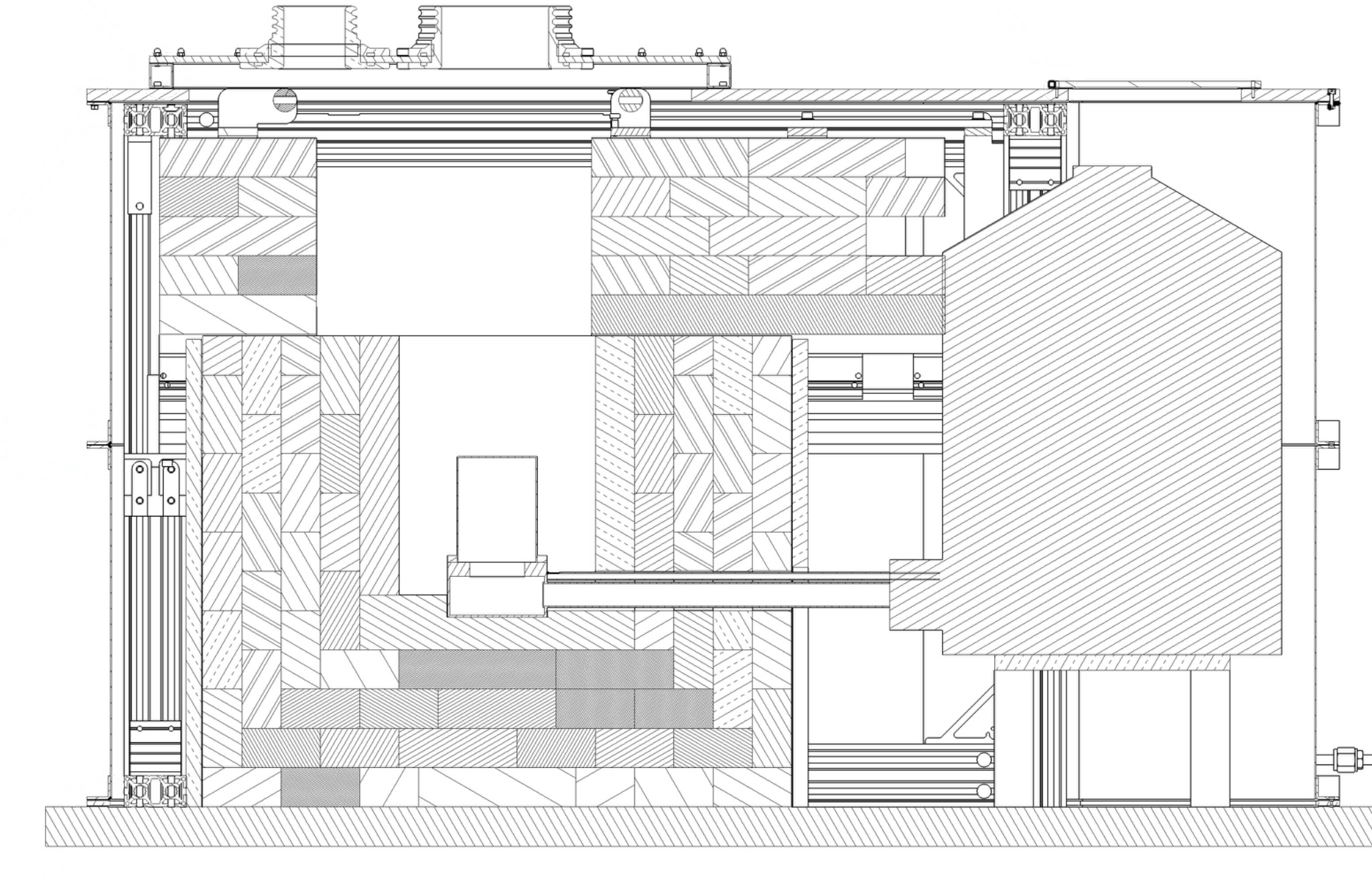}
\caption{Section view of the Gator HPGe facility. The copper cryostat is located
inside a copper-lined sample cavity, surrounded by lead and polyethylene
shielding within a stainless-steel enclosure.  A U-shaped
copper cold finger immersed in liquid nitrogen provides the cooling of the crystal.}
\label{fig:facility}
\end{figure}

The sample cavity measures $25\times25\times33$\,cm$^3$ and is surrounded by
oxygen-free high-conductivity copper, with a thickness of 5\,cm on the sides
and 7\,cm at the base. The copper is enclosed by a 5\,cm inner lead layer and
15\,cm of outer lead, with nominal $^{210}$Pb activities of 3 and
75\,Bq/kg, respectively. A 5\,cm layer of borated polyethylene provides
additional shielding against ambient neutrons. The facility layout is shown schematically 
in Fig.~\ref{fig:facility}.

A stainless-steel enclosure and lateral load-lock chamber were introduced
in the earlier upgrade described in Ref.~\cite{Gator2022}. Samples can be
purged with gaseous nitrogen in the load-lock before transfer to the cavity,
which is maintained under a continuous nitrogen purge at positive pressure
to suppress radon ingress. Glove ports in an acrylic plate above the cavity
allow for sample handling within the enclosure. 

\subsection{Motivation and hardware changes}
\label{sec:upgrade}

The earlier readout used an ORTEC 672 spectroscopy amplifier followed by an ORTEC ASPEC-927 multichannel analyser (MCA), which stored pulse-height spectra~\cite{Gator2011,Gator2024}. Individual waveforms were thus not retained, limiting the information available for analysis, and for investigating electronic artefacts.
Broad noise-related features below approximately 100\,keV, as well as increased analysis flexibility, motivated the replacement of readout components and the introduction of waveform acquisition.

\begin{figure}[tbp]
\centering
\includegraphics[width=\columnwidth]{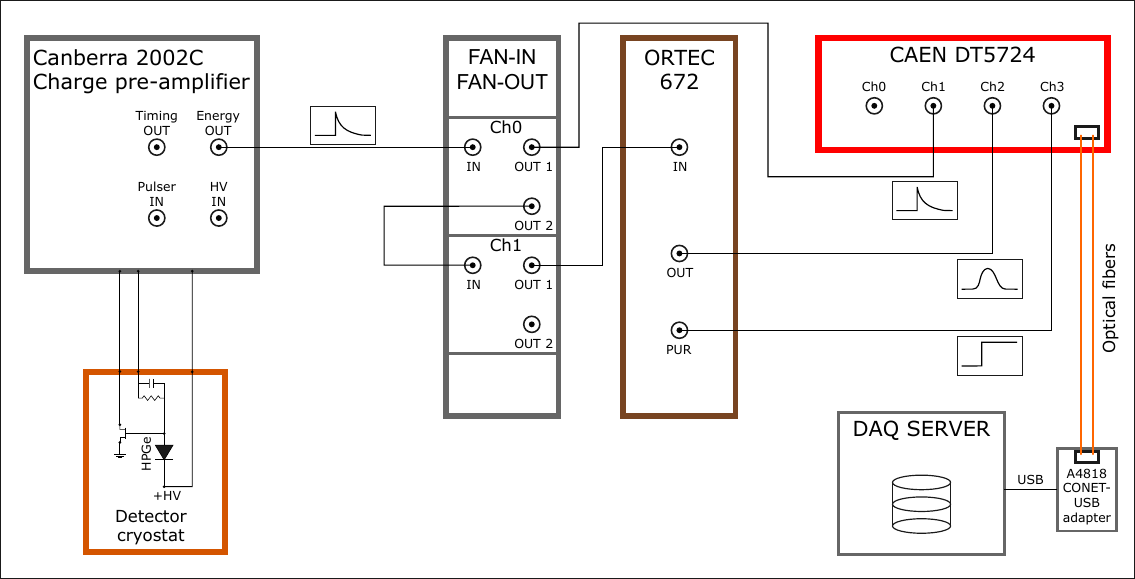}
\caption{ 
Signal routing in the upgraded readout system, showing the devices relevant to signal acquisition and distribution. Only the inputs and outputs involved in the signal routing are represented. The HPGe detector is schematically represented as a reverse-biased diode inside the cryostat (orange box), with the preamplifier input JFET and feedback RC network located close to the crystal. The HV input of the Canberra 2002C is omitted for clarity. Data readout is performed via an A4818 CONET-USB adapter connected to the digitiser through optical fibres.}
\label{fig:readout}
\end{figure}

The upgrade replaces the ASPEC-927 MCA with a CAEN DT5724 digitiser, while retaining the ORTEC 672 spectroscopy amplifier.  A custom, low-noise NIM fan-out module, constructed by the LNGS Electronics Service, splits the preamplifier signal for digitisation and shaping. Its measured bandwidth and noise spectral density are 100\,MHz and $~1.6\,\mathrm{nV}/\sqrt{\mathrm{Hz}}$, respectively. The signal cables were replaced with low-noise cables using MCX and SMA connectors. Figure~\ref{fig:readout} shows the  signal-routing diagram.

\subsection{Waveform acquisition and energy reconstruction}
\label{sec:daq}

The output of the Canberra 2002C charge-sensitive preamplifier is split by a custom fan-in/fan-out module. One branch is connected directly to a digitiser input, while the other is fed into the ORTEC 672 spectroscopy amplifier, which shapes the preamplifier signal and provides a pile-up rejection (PUR) signal. The upgraded DAQ system simultaneously records the preamplifier output, the shaped output of the spectroscopy amplifier, and its PUR signal.
Simultaneous acquisition allows the preamplifier pulses to be compared with their corresponding shaped waveforms, providing information for the identification of pile-up events and other noise sources. 

The three signals are digitised by the CAEN DT5724 at a sampling rate of {100\,MS/s}, with {14-bit} resolution and a {2.25\,V} unipolar input range. Event acquisition relies on the digitiser's channel self-triggering functionality, configured on the spectroscopy amplifier signal. The threshold is set to 45 ADC counts above the baseline, equivalent to approximately {6.2\,mV} for the nominal input range.
Like the previous MCA-based DAQ, the upgraded system triggers on the spectroscopy-pulse amplitude, which is related to the collected charge. The threshold of the new system is low enough to support a reasonably fair comparison of the low-energy region of the two background spectra, where the previous system showed a broad feature attributed to electromagnetic interference.

Waveform storage allows for the baseline properties to be evaluated for individual events and used in the data quality selection described in Sect.~\ref{sec:selection}. Energy reconstruction is performed offline, with adjustable shaping and parameters. 
The energy scale and resolution are determined
from calibration measurements as described in Sect.~\ref{sec:calibration}.

\section{Data sets and analysis methods}
\label{sec:methods}

In this section we discuss the characterisation of the detector after
the readout upgrade. We describe the acquired data sets and the new
analysis procedures, while the performance of the detector is presented in
Sect.~\ref{sec:performance}.

\subsection{Measurements}
\label{sec:datasets}

The performance study uses a dedicated background measurement after the
upgrade and measurements of a granite sample taken from tunnel meter 3500 of the Bedretto underground
site~\cite{bedretto_charaterization,Bedretto2026} before and after the readout changes. The granite spectrum contains
lines from the primordial uranium and thorium decay chains and from $^{40}$K, providing
peaks across the energy range used to calibrate the energy scale and characterise the energy resolution.
The post-upgrade background measurement is also compared with an earlier
background spectrum to assess changes in the noise conditions, 
continuum and identifiable $\gamma$-lines.

The PMT assay comprises a measurement of sixteen tubes in PTFE holders and
a separate background measurement with holders but without the PMTs. The PMT 
sample measurement, which was carried out before the upgrade, has a live time of 102.2\,days.
The holder background run, with a live time of 57.5\,d, constrains contributions 
from the detector environment and the support material in the activity analysis, 
and it is distinct from the background run used to characterise the upgraded facility. 

\subsection{Data processing and data quality selection}
\label{sec:selection}

Data are acquired using custom C++ DAQ software developed for flexible configuration of CAEN digitizers and publicly available in a GitHub repository~\cite{fmcdaqrepo}.
Digitizer and acquisition settings, including register programming, are specified in a JSON configuration file.
Each event comprises three digitized waveforms, each $70\,\mu\mathrm{s}$ long (7000 samples) with a $15\,\mu\mathrm{s}$ pre-trigger interval (1500 samples).
The signals have positive polarity; the digitizer input offset is chosen such that 98\% of the full dynamic range can be used, 
while retaining sufficient range for small negative-going signal excursions.
Waveforms are stored as arrays in a ROOT tree. Files are written at approximately two-hour intervals.
The acquisition configuration, digitizer metadata, and each file's start and stop timestamps are stored for offline processing and live-time determination.
Each event also includes a $20\,\mathrm{ns}$ resolved event time stamp.

The raw ROOT files are processed offline by a waveform processor that is part of a publicly available repository supporting the automated DAQ data transfer and processing pipeline~\cite{gatortoolsrepo}. 
Processing is configured through a JSON file with channel-specific settings that define the operations applied to each waveform and the output observables. 
The extracted event quantities are saved in a dataframe in NumPy format.

For the preamplifier and spectroscopy-amplifier signals, the extracted quantities include the baseline mean, RMS, median and median absolute deviation. 
The preamplifier waveform is processed with a trapezoidal filter with a shaping time of $15\,\mu\mathrm{s}$ and a flat top of $0.8\,\mu\mathrm{s}$. 
Its decay-time parameter, $\tau=46.6\,\mu\mathrm{s}$, was determined by fitting an exponential to the averaged preamplifier waveform in a dedicated test-pulser dataset, in which the pulser signal was capacitively coupled to the preamplifier input. The shaped spectroscopy-amplifier waveform is smoothed with a Gaussian filter using $\sigma=75$ samples and a kernel half-width of 40 samples (81 samples across the full kernel). 
The maximum of the smoothed pulse provides an event-by-event pulse-height estimate analogous to the previous MCA measurement, without binning the values into a spectrum. 
The PUR signal is processed separately to flag pile-up events.

\begin{figure*}[tbp]
\centering
\includegraphics[width=0.48\textwidth]{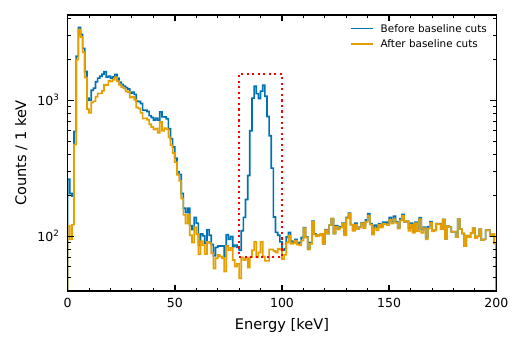}
\hfill
\includegraphics[width=0.48\textwidth]{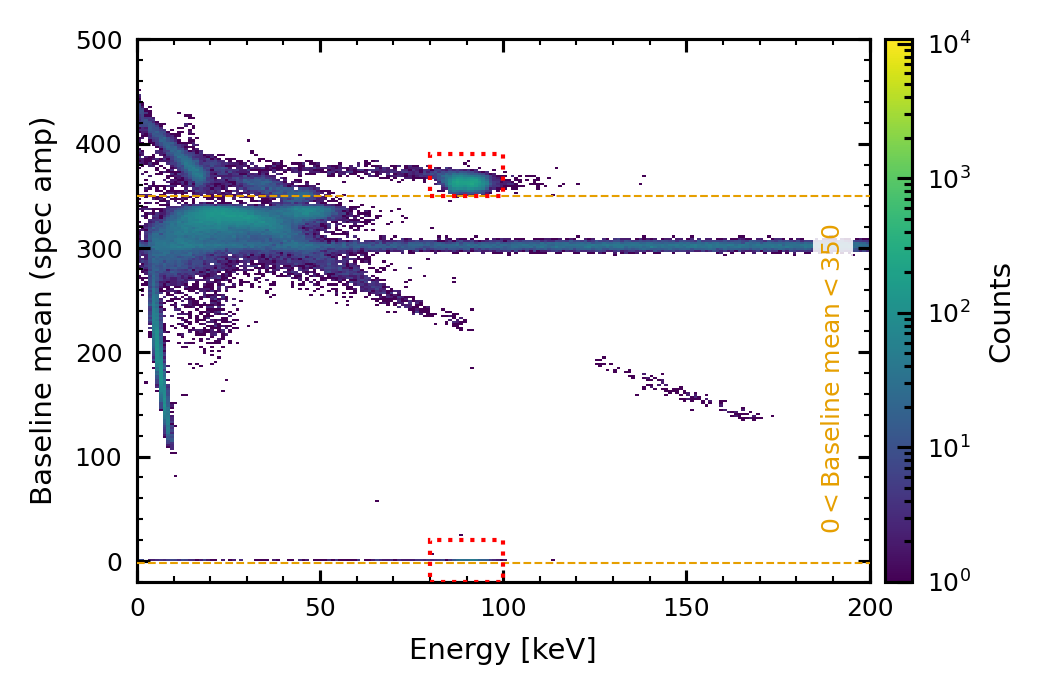}
\caption{(Left): Energy spectrum in the (0-200)\,keV energy range before (blue) and after (orange) the baseline cut. The artefact peak is visible just below 100\,keV. (Right): Baseline of the spectroscopy amplifier waveform versus reconstructed energy. A population of events (marked in red) corresponding to the peak artefact can be seen, along with a distribution of events at zero baseline. These two populations are removed by the baseline cut. The baseline mean selection is shown by the orange dashed line. The colour scale gives the number of events.}
\label{fig:baseline}
\end{figure*}

We use the new waveform information to identify electronic artifacts in the post-upgrade background data. Figure~\ref{fig:baseline} shows the baseline
mean as a function of reconstructed energy for the preamplifier and
spectroscopy-amplifier signals. In addition to the main population, the
distributions contain events with anomalous baseline values, including a
population near zero. A selection on the baseline mean removes the
noise-related structure near 100\,keV in the original spectrum. The selected
background spectrum is discussed in Sect.~\ref{sec:backgrounds}.

\subsection{Energy calibration and peak modeling}
\label{sec:calibration}

The energy resolution comparison is based on fits to individual
$\gamma$-ray peaks in the Bedretto granite spectra, acquired before and after the upgrade. 
Each peak is described by a Crystal Ball function and a sigmoid step background. 
The Crystal Ball function combines a Gaussian core with a low-energy power-law tail to
describe departures from a symmetric peak shape, following the approach we used previously for Gator calibrations. The peak widths are expressed as full widths at half maximum (FWHM) and compared as a function of energy in Sect.~\ref{sec:resolution}.

\begin{figure}[!h]
\centering
\includegraphics[width=\columnwidth]{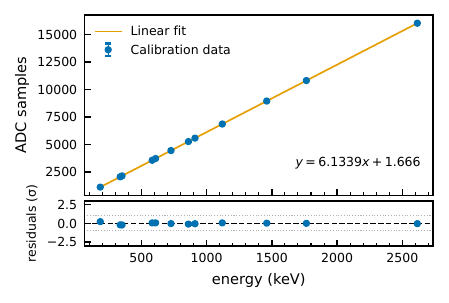}
\caption{Energy calibration performed after the upgrade, using the Bedretto granite sample.}
\label{fig:cal_post_upgrade}
\end{figure}

\subsection{Detection efficiency simulations}
\label{sec:efficiency}

We obtained the detection efficiencies for the PMT samples 
with Geant4-based  Monte Carlo simulations\,\cite{Agostinelli:2002hh} of their geometry in the Gator cavity. The physical arrangement and the simulation model are presented in Sect.~\ref{sec:pmt_setup}.
The efficiency depends on the photon energy, the positions of the radioactive
emitters, and the attenuation in the sample and intervening materials.

The Gator detector simulation framework was previously tested with calibrated point sources,
as well as with extended sources; see~\cite{Gator2011,Gator2022}. Earlier PMT studies also
demonstrated the dependence of the inferred activity on the assumed spatial
distribution of impurities, as we showed in Ref.~\cite{Gator2022}. 
The simulations were performed with Geant4. The PMT model consists of a Kovar envelope enclosing a vacuum volume, which contains a 3.1\,mm thick SiO$_2$ window at the front face, a stem at the back, and a homogenised cylinder representing the internal electrode structure, for a total mass of $\approx$183\,g per PMT. 
The stems are modelled as a cobalt-free metal disk in which each of the sixteen feedthrough pins is sealed by an individual borosilicate glass spot, replacing the ceramic ring of the standard design (see Figure~\ref{fig:pmt_stem_improvement}).
The PTFE holders were included in the geometry as passive material. For each isotope, $10^7$ decays were simulated assuming uniform distribution in the materials of the sixteen PMTs. The branching ratios are taken from Laboratoire National Henri Becquerel tables of recommended data\,\cite{TabRad_v4}. For the $^{208}$Tl they include the $^{212}$Bi$\to^{208}$Tl branching fraction of 35.94\%. We list the line efficiencies
and branching ratios for the sixteen PMT geometries in Table~\ref{tab:pmt_efficiencies} in Appendix\,\ref{app:det_efficiencies}.

\subsection{Activity estimation and upper limits}
\label{sec:statistics}

We compare two approaches for the PMT assay: a peak-counting analysis and a
joint Poisson likelihood analysis. In both approaches, sample and background
measurements constrain the contribution of the PMT units to the lines of interest.
We report activities per PMT for individual radionuclides or decay-chain segments. 

In the counting analysis, the continuum underlying a peak is estimated from
adjacent sidebands. If $S_k$ and $B_k$ denote the continuum-subtracted peak
counts for line $k$ in the sample ($S$) and background ($B$) measurements, respectively,
the net sample contribution is:
\begin{equation}
S_{\mathrm{net},k}=S_k-\frac{t_S}{t_B}B_k,
\label{eq:net_counts}
\end{equation}
where $t_S$ and $t_B$ are the respective live times of the signal and background measurement. This notation follows
our previous Gator counting analysis~\cite{Gator2022}. The net counts are
then converted to an activity using the line branching ratio, simulated detection
efficiency, measurement live time and number of PMTs. An activity or an upper limit is reported if $S_{net}$ exceeds the detection limit $L_d$ as defined in\,\cite{Hurtgen2000}. If several lines of a given radionuclide are detected, the activity is calculated as the error-weighted average. If none is detected, the lowest of the individual upper limits is adopted.

The likelihood approach uses the observed counts directly, retaining
separate Poisson terms for the peak and sideband regions of the sample and background spectra. For each line $k$, the peak region is defined as $\pm3\sigma_k$ around the nominal line energy, where
$\sigma_k$ denotes the corresponding energy resolution. The sideband region consists of two adjacent windows of width $3\sigma_k$, one on each side of the peak, summed into a single region. Counting the events in the peak and sideband regions of both the sample and background spectra gives four observed counts per line, $n_{ik}$
with $i=1,\dots,4$, corresponding to the sample peak, sample sideband, background peak, and background sideband, respectively.
The log-likelihood has the following form:

\begin{equation}
 \ln\mathcal{L}=\sum_k\sum_{i=1}^{4}
 \ln\mathrm{Pois}(n_{ik}\mid\lambda_{ik}),
 \label{eq:joint_likelihood}
\end{equation}
where $\lambda_{ik}$ are the expected counts, defined for each region as
\begin{align}
 \lambda_{1k} &= t_S\,(s\,\epsilon_k r_k+\beta+c+b), &
 \lambda_{2k} &= t_S\,\rho_k\,(c+b), \nonumber\\
 \lambda_{3k} &= t_B\,(\beta+b), &
 \lambda_{4k} &= t_B\,\rho_k\,b.
 \label{eq:expectations}
\end{align}
Here $t_s$ and $t_b$ are the live times of the sample and background measurements, respectively, $\epsilon_k$ and $r_k$ are the detection efficiency and the branching ratio of line $k$, and $\rho_k$ is the ratio of the total sideband width to the peak-window width ($\rho_k=1$ here). In this notation, $s$ denotes the sample signal normalisation before multiplication by the line efficiency and branching ratio, $\beta$ is the background peak contribution, $c$ the additional sample continuum, and $b$ the flat background. The parameter $s$, is shared by all lines of the same radionuclide or decay-chain segment, assuming secular equilibrium within the segment. The parameters $\beta$, $c$ and $b$ are nuisance parameters, fitted independently for each line to account for the energy dependence of the background components, and constrained only to be non-negative.
The best-fit activity is found by minimising $-2\ln\mathcal{L}$ with \texttt{iminuit}\,\cite{iminuit}. Confidence intervals are derived from the profile likelihood, as the range of $s$ for which $-2\ln\mathcal{L}$ exceeds its minimum by less than 2.71. A radionuclide is considered detected if the lower bound of this interval excludes $s=0$; otherwise, its upper edge is quoted as a one-sided upper limit at 95\% confidence level.
The uncertainty on the best-fit activity is given by the quadrature sum of the statistical uncertainty, obtained from the profile likelihood, and a 10\% systematic uncertainty on the efficiency. The latter is estimated by comparing simulated efficiencies to measurements of certified reference sources\,\cite{Gator2011}.

The counting and likelihood analyses provide a comparison of the activity
estimates obtained from the same PMT sample. Hence, with this comparison we test the
dependence of the results on the statistical treatment of the peak and background counts. The results from the 
two methods are compared in Sect.~\ref{sec:pmt_results}.

\section{Detector performance after the upgrade}
\label{sec:performance}

We assess the performance after the readout upgrade from the resolution
of the granite calibration peaks and the selected background spectrum.
The sensitivity study combines these measurements with the simulated
efficiencies for the 3-inch PMT sample. 

\subsection{Energy resolution}
\label{sec:resolution}

\begin{figure}[tbp]
\centering
\includegraphics[width=\columnwidth]{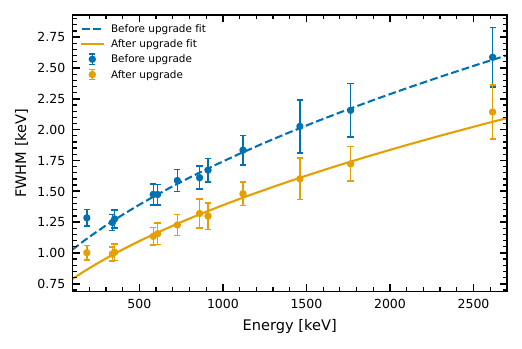}
\caption{Energy resolution from the Bedretto granite measurements before
(blue points) and after (orange points) the readout upgrade. The data points show
the fitted resolutions (at FWHM) of the various peaks, with their uncertainties. The two curves show the
corresponding fits to the data, before (dotted) and after (solid) the upgrade.}
\label{fig:resolution_comparison}
\end{figure}

Figure~\ref{fig:resolution_comparison} compares the peak widths obtained
from the granite measurements before and after the upgrade. The
$\gamma$-peaks in the post-upgrade spectrum have smaller resolution values over the measured energy range,
including the $^{208}$Tl line at 2614.5\,keV. A higher than expected resolution is observed at an energy of 186\,keV. This is caused by the overlap of two $\gamma$-lines from $^{226}$Ra (186.21\,keV) and $^{235}$U (185.72\,keV) which cannot be resolved individually.

The two resolution curves are used in the
sensitivity comparison presented in Sect.~\ref{sec:sensitivity}.
In Table \ref{tab:old_new_fwhm} we also show the energy resolution of the major $\gamma$-lines before and after upgrade.

\begin{table}[tbp]
\caption{Energy resolution (at FWHM) for the major $\gamma$-lines in the (100-2700)\,keV energy region, before and after the DAQ upgrade.}
\label{tab:old_new_fwhm}
\centering
\small
\begin{threeparttable}
\begin{tabular}{rlrr}
\hline\noalign{\smallskip}
Energy & \multirow{2}{*}{Nuclide} &  \multicolumn{2}{c}{FWHM [keV]} \\
\multicolumn{1}{c}{[keV]}   &  & \multicolumn{1}{c}{Old}      &   \multicolumn{1}{c}{New} \\
\noalign{\smallskip}\hline\noalign{\smallskip}
185.72   & $^{235}$U\,\tnote{*}  & $1.28 \pm 0.07$    & $1.00 \pm 0.06$ \\
338.32   & $^{228}$Ac & $1.25 \pm 0.06$    & $0.99 \pm 0.06$ \\
351.93  & $^{214}$Pb & $1.28 \pm 0.07$    & $1.01 \pm 0.07$ \\
583.19  & $^{208}$Tl & $1.47 \pm 0.08$   & $1.13 \pm 0.07$ \\
609.31  & $^{214}$Bi & $1.47 \pm 0.08$   & $1.16 \pm 0.09$ \\
727.33  & $^{212}$Bi & $1.59 \pm 0.09$   & $1.23 \pm 0.08$  \\
860.56  & $^{208}$Tl & $1.61 \pm 0.09$   & $1.32 \pm 0.12$ \\
911.20  & $^{228}$Ac & $1.67 \pm 0.10$    & $1.30 \pm 0.11$ \\
1120.29  & $^{214}$Bi & $1.83 \pm 0.12$   & $1.48 \pm 0.10$ \\
1460.88  & $^{40}$K   & $2.02 \pm 0.21$   & $1.60 \pm 0.17$ \\
1764.49  & $^{214}$Bi & $2.16 \pm 0.22$   & $1.72 \pm 0.14$ \\
2614.51  & $^{208}$Tl & $2.59 \pm 0.24$   & $2.14 \pm 0.22$ \\
\noalign{\smallskip}\hline
\end{tabular}
\begin{tablenotes}
\item[*] Combined line including the 186.72\,keV emission of $^{226}$Ra
\end{tablenotes}
\end{threeparttable}
\end{table}

\subsection{Background spectrum and comparison with earlier measurements}
\label{sec:backgrounds}

\begin{figure*}[tbp]
\centering
\includegraphics[width=\textwidth]{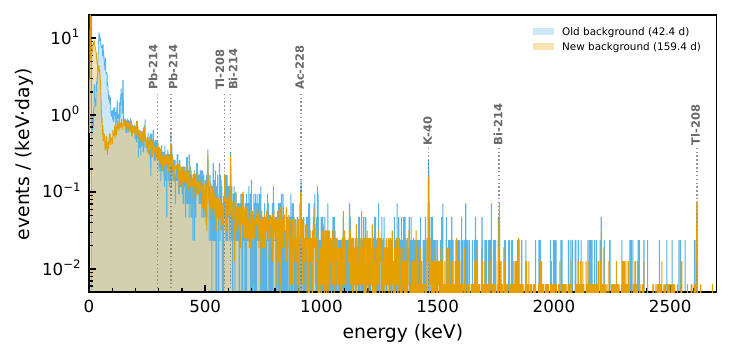}
\caption{Background spectra before (orange) and after (blue) the readout
upgrade, normalised per unit energy and measurement time. The post-upgrade
spectrum includes the baseline selection. The measurement durations were 
42.4 and 159.4\,days, respectively.}
\label{fig:background_comparison}
\end{figure*}

We compare the background spectra before and after the upgrade in
Fig.~\ref{fig:background_comparison}. The post-upgrade spectrum includes
the baseline selection described in Sect.~\ref{sec:selection}. The
noise-related structure near 100\,keV,  visible in the earlier spectrum, is
suppressed in the data after applying the event selection criteria. 

The comparison of the integrated rates in the (100--2700)\,keV energy region shows
a reduction of about 9\% relative to the pre-upgrade background dataset. 
We distinguish this comparison from our background measurement
reported in Ref.~\cite{Gator2022}, for which the integrated rate in the
same energy interval was $(82.0\pm0.7)$\,events/(kg\,day). The present
background level does not show a reduction relative to this earlier published value. The observed excess of events below $\sim$20\,keV remains under investigation, 
and we will discuss further measures in Sect.~\ref{sec:conclusions}.

\subsection{Sensitivity for a specified sample geometry}
\label{sec:sensitivity}

\begin{figure}[tbp]
\centering
\includegraphics[width=0.97\columnwidth]{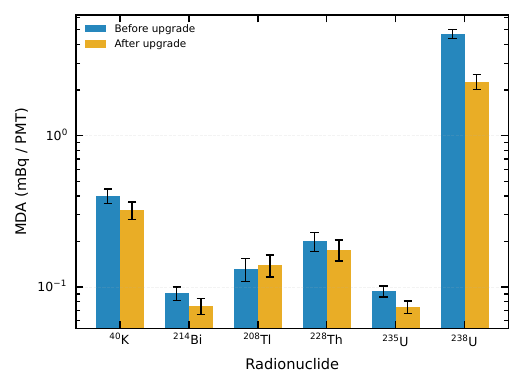}
\caption{Projected Minimum Detectable Activity (MDA), in mBq/PMT  at 95\% CL, for prominent radionuclides, before and after the DAQ upgrade. A measurement time of 42 days is assumed.}
\label{fig:k40_sensitivity}
\end{figure}

We evaluated the spectrometer's sensitivity using the geometry of the low-activity stem,
3-inch PMT sample (see Figure\,\ref{fig:pmt_geometry}). To this end, we used the simulated detection efficiencies and the
measured background in windows around selected $\gamma$-ray lines. The study considers the $^{214}$Bi line at 609.3\,keV, the $^{228}$Ac line at
911.2\,keV, the $^{40}$K line at 1460.8\,keV and the $^{208}$Tl line at 2614.5\,keV. 
The counting window extends to $\pm1.5\sigma$ around each $\gamma$-line, corresponding to a containment fraction of $\sim$0.866 for a Gaussian peak.
We then obtain the Minimum Detectable Activity (MDA) by dividing the detection limit in counts by the
product of counting time, line efficiency, branching ratio and peak
containment. The efficiency is estimated for the entire batch of 16 PMTs. Thus, the MDA calculated here represents the sensitivity of the detector to the PMT batch measurement. To obtain the MDA per PMT, as shown in Figure\,\ref{fig:k40_sensitivity}, the activity is divided by 16.

The comparison shows that the
sensitivity change depends on the selected line. Among the selected lines, sensitivity improves for the low-energy lines of $^{238}$U (92.6 keV) and $^{235}$U (185.7 keV); for $^{40}$K (1460.8 keV), $^{214}$Bi (609.3 keV), $^{208}$Tl (2614.5 keV) and $^{228}$Th (911.2 keV), the sensitivities measured before and after the upgrade are consistent within $1\sigma$. The sensitivity uncertainty includes a 5\% systematic contribution from GEANT4 cross-section models, propagated through the photoabsorption-efficiency estimate\,\cite{Cirrone:2010zz}, and a contribution from background-counting statistics.

\section{Radioassay of low-activity-stem PMTs}
\label{sec:samples}

\begin{figure}
    \centering
    \includegraphics[width=0.97\linewidth]{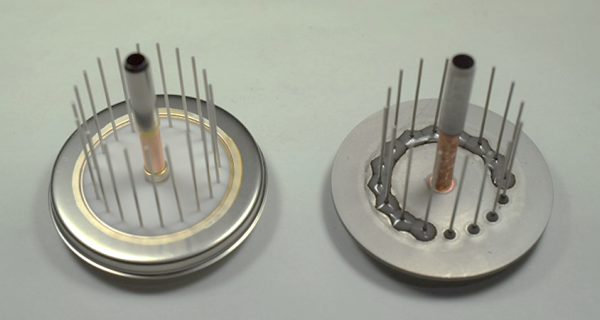}
    \caption{Improvements in the stem design at the back of the R17317 PMTs. The original design (left) using a large ceramic plate for the stem is improved using low activity cobalt-free metal and smaller borosilicate glass beads (right).}
    \label{fig:pmt_stem_improvement}
\end{figure}

To provide the internal components of the PMT such as the photocathode, the dynode chain and the screening electrode need to be delivered with high voltage. This is applied using pins protruding from the back side of the vacuum tube and being sealed with a stem. In the previous model of the R17317 PMTs this seal is established via a ceramic plate (Figure\,\ref{fig:pmt_stem_improvement}, left). Since the ceramic stem contributes a significant fraction to the overall radio isotope activity, a new stem design (Figure\,\ref{fig:pmt_stem_improvement}, right) has been developed. The new design replaces the ceramic disk with cobalt-free metal, leaving only a small amount of borosilicate glass beads sealing the pin feedthrough. Two iterations of the new stem design have been screened individually (excluding the multiplier tube) beforehand using Gator with results reported in\,\cite{Bismark:2024vyl}.

\subsection{Sample description and measurement geometry}
\label{sec:pmt_setup}

\begin{figure}
    \centering
    \includegraphics[width=0.8\linewidth]{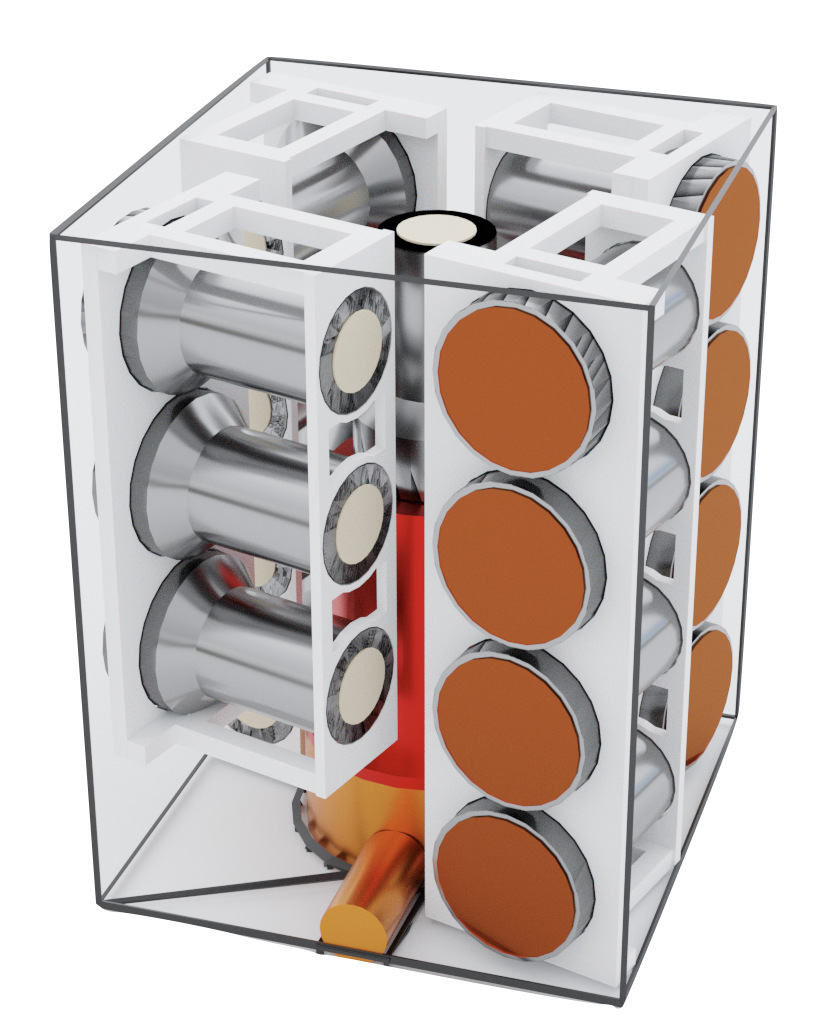}
    \caption{Arrangement of the sixteen R17317 PMTs in the Gator cavity as implemented in the geometry used for the detection efficiency MC simulations. Fifteen tubes are supported in PTFE holders (white) around the copper cryostat (red), while one is placed on the cryostat endcap.}
    \label{fig:pmt_geometry}
\end{figure}

The sample consists of sixteen Hamamatsu R17317 3-inch PMTs with
low-activity stems, investigated as candidate photosensors for the XLZD
TPC~\cite{XLZDDesignBook}.
Fifteen tubes were positioned around the detector in four PTFE holders:
three holders accommodated four tubes each, while only three tubes could be accommodated in the fourth holder due to the presence of the HPGe crystal's cryostat elbow (see Figure\,\ref{fig:facility}). The remaining tube was placed on the cryostat endcap. In Figure~\ref{fig:pmt_geometry} we show the arrangement and its implementation
in the efficiency MC simulation. The sample measurement duration is
approximately 102.2\,days, as introduced in Sect.~\ref{sec:datasets}.

Separate measurements with the PTFE holders and without PMTs provide the
background for the assay. These include configurations with three and two holders. 

\subsection{Measured spectra and activities}
\label{sec:pmt_results}

\begin{figure*}[tbp]
\centering
\includegraphics[width=\textwidth]{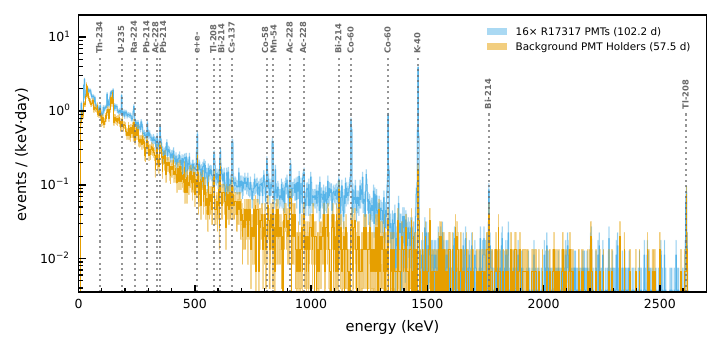}
\caption{Measured energy spectra of the sixteen PMTs with holders (blue) and the holder background (orange).
The durations of the two runs was 102.17 and 57.46\,days, respectively. Statistical uncertainties are indicated as transparent bands.}
\label{fig:pmt_spectrum}
\end{figure*}

We show the sample and PTFE holder background spectra in
Figure~\ref{fig:pmt_spectrum}. The sample contributes both continuum events
and identifiable peaks above the holder background. As examples, the $^{40}$K peak and
the two $^{60}$Co peaks are prominent, as well as lines from the uranium and thorium
decay chains. The likelihood analysis uses the sample and background counts separately,
as we described in Sect.~\ref{sec:statistics}. The obtained activities or upper limits are shown in 
Table~\ref{tab:pmt_methods}.  While the $^{238}$U contribution is constrained by an upper limit, 
we obtain  finite activities for all the other inspected nuclides. 

We also show the comparison between the counting and likelihood activity estimates  in
Table~\ref{tab:pmt_methods}. In the counting analysis, we obtain upper limits
for $^{226}$Ra, $^{228}$Ra and $^{228}$Th, while the likelihood analysis
returns finite values for these chain segments. For the nuclides
with finite activities  in both methods, the central values are close, and agree within 
uncertainties. 

\begin{table}[tbp]
\caption{Comparison of the counting and likelihood analyses
of the sixteen PMT sample. Entries are activities or upper limits in
mBq/PMT. The upper limits are quoted at 95\% confidence level.}
\label{tab:pmt_methods}
\centering
\begin{tabular}{lcc}
\hline\noalign{\smallskip}
Nuclide  & Counting & Likelihood  \\
\noalign{\smallskip}\hline\noalign{\smallskip}
$^{238}$U  & $<6.4$           & $<3.5$  \\
$^{226}$Ra & $<0.15$        & $0.13\pm0.04$  \\
$^{228}$Ra & $<0.29$        & $0.12\pm0.05$ \\
$^{228}$Th & $<0.18$        & $0.12\pm0.05$ \\
$^{235}$U  & $0.18\pm0.04$  & $0.18\pm0.03$ \\
$^{40}$K   & $12.0\pm1.3$     & $12.0\pm1.3$ \\
$^{60}$Co  & $0.30\pm0.03$  & $0.29\pm0.03$ \\
$^{137}$Cs & $0.12\pm0.02$  & $0.12\pm0.02$ \\
$^{54}$Mn & $0.11\pm0.02$  & $0.11\pm0.02$ \\
$^{58}$Co & $0.04\pm0.01$  & $0.03\pm0.01$ \\
\noalign{\smallskip}\hline
\end{tabular}
\end{table}

In Table\,\ref{tab:pmt_xenonnt}, we finally compare the results of the PMTs with low radioactivity stems with those from the 3-inch PMTs used in the ongoing XENONnT~\cite{XENON:2021mrg}, LZ~\cite{LZRadiopurity2020} and PandaX-4T~\cite{PandaXRadiopurity2022} experiments. 
We find  lower central activities for $^{226}$Ra, $^{228}$Ra, $^{228}$Th and $^{60}$Co than the
XENONnT and LZ values. These are promising results for XLZD, which aims for a further background reduction compared to the existing two-phase xenon TPCs.

\begin{table}[tbp]
\caption{Comparison of activities in mBq/PMT for the 16 R17317 sample and the 3-inch PMTs screened for XENONnT, LZ and PandaX-4T. The XENONnT activities are 
averages for 180 tubes~\cite{XENON:2021mrg}. The LZ values
are averages from 230 tubes~\cite{LZRadiopurity2020}. The PandaX-4T values are averages from 368 tubes\,\cite{PandaXRadiopurity2022}.}
\label{tab:pmt_xenonnt}
\centering
\small
\setlength{\tabcolsep}{4pt}
\begin{threeparttable}
\begin{tabular}{lcccc}
\hline\noalign{\smallskip}
Nuclide & R17317 & XENONnT & LZ\,\tnote{\textsuperscript{\textdagger}} & PandaX-4T\,\tnote{\textsuperscript{\textdagger}}\\
\noalign{\smallskip}\hline\noalign{\smallskip}
$^{238}$U  & $<3.5$         & $9\pm2$        & $5.0\pm2.1$ & $<56.5$ \\
$^{226}$Ra & $0.13\pm0.04$  & $0.47\pm0.02$  & $0.9\pm0.2$ & $<3.99$\\
$^{228}$Ra & $0.12\pm0.05$  & $0.47\pm0.07$  & $1.4\pm0.8$ & $<7.82$\\
$^{228}$Th & $0.12\pm0.05$  & $0.46\pm0.02$  & $0.8\pm0.2$ & $<3.06$\\
$^{235}$U  & $0.18\pm0.03$  & $0.40\pm0.10$  & $0.8\pm0.6$ & $<28.3$\\
$^{60}$Co  & $0.29\pm0.03$  & $1.05\pm0.03$  & $1.9\pm0.2$ & $<2.34$\\
$^{40}$K   & $12.0\pm1.3$   & $14.2\pm0.5$   & $12.2\pm1.1$ & $<22.3$\\
$^{137}$Cs & $0.12\pm0.02$  & $<0.14$        & $0.51\pm0.03$ & $<1.85$ \\
\noalign{\smallskip}\hline
\end{tabular}
\begin{tablenotes}
\item[\textsuperscript{\textdagger}] Reported upper limits at 90\% confidence level.
\end{tablenotes}
\end{threeparttable}
\end{table}

\section{Conclusions and outlook}
\label{sec:conclusions}

We have described an upgrade of the Gator readout chain, that introduces waveform
digitisation along with changes to the spectroscopy electronics, signal
distribution and cabling. The recording of the preamplifier and amplifier signals
provides information for event selection and offline energy reconstruction
that was not retained in the earlier MCA spectra. A baseline selection
suppresses pick-up noise around $\sim$100\,keV in the background data acquired after the upgrade.

Measurements of a granite sample from the Bedretto laboratory\,\cite{Bedretto2026} show narrower $\gamma$-ray
peaks after the upgrade, with an improvement of energy resolution of 17\% at 2614.5\,keV. Our 
background comparison yield a 9\% reduction in the
integrated rate between 100 and 2700\,keV relative to our pre-upgrade measurement.  Due to the improved energy resolution and lower background rate,  the sensitivity of the 
spectrometer (for our specified PMT geometry) shows improvements for several lines, especially for the ones at low energies.

The measurement of sixteen Hamamatsu R17317 PMTs with low-activity stems yields activities for 
individual radionuclides and decay-chain segments, together with an upper limit for $^{238}$U. 
Our comparison with the XENONnT and LZ PMT screening results show lower central activities for
$^{226}$Ra, $^{228}$Ra, $^{228}$Th and $^{60}$Co.  While the concrete implications for the XLZD background budget 
require dedicated Monte Carlo simulations, these lower activities are central for the science goals of the 
observatory, in particular also for the neutrinoless double beta decay channel~\cite{XLZD:2024pdv}.

Further characterisation of the Gator facility will address the noise below 20\,keV including the possibility of its further suppression using the now available option of pulse shape analysis. Hardware upgrades planned for the coming year will include isolating the LN$_2$ refill line, improving the electrical feedthroughs and signal transmission lines for improved control of electromagnetic interference.

The facility's planned screening programme includes candidate materials for the future XLZD, LEGEND-1000 and QROCODILE~\cite{QROCODILE:2024nqm} experiments. 

\begin{acknowledgements}
We acknowledge support from the Swiss National Foundation via the ongoing grants 20FL20-232696, 20FL20-232670,  2000-1-240028 and 2000-1-243007. We thank the University of Zurich for continued support, as well as the mechanical and electronics workshop of the Physics Department.  We thank 
the electronics workshop at LNGS for the custom fan-in/fan-out module. We thank Knut Dundas Mor{\aa} for helpful discussions. We also thank Hamamatsu Photonics for providing the 3-inch PMTs and PMT stem samples.
\end{acknowledgements}

\pagebreak
\begin{appendices}
\section{Detection efficiencies}\label{app:det_efficiencies}

\vspace{-2\baselineskip}
\begin{table}[h]
\caption{Detection efficiencies and branching ratios (BR) used for the sixteen-PMT sample measurement.}
\label{tab:pmt_efficiencies}
\centering
\small
\begin{tabular}{SlSS}
\hline\noalign{\smallskip}
\multicolumn{1}{l}{Energy [keV]} & Nuclide & \multicolumn{1}{l}{BR} & \multicolumn{1}{l}{$\epsilon~[\%]$} \\
\noalign{\smallskip}\hline\noalign{\smallskip}
92.6     & $^{234}$Th & 0.043   & 0.466 \\
185.72   & $^{235}$U  & 0.572    & 1.34 \\
238.632  & $^{212}$Pb & 0.436    & 1.15 \\
295.224  & $^{214}$Pb & 0.184    & 1.17 \\
338.32   & $^{228}$Ac & 0.114    & 1.15 \\
351.932  & $^{214}$Pb & 0.356    & 1.14 \\
583.187  & $^{208}$Tl & 0.305   & 0.798 \\
609.312  & $^{214}$Bi & 0.455   & 0.837 \\
661.657  & $^{137}$Cs & 0.849   & 0.915 \\
810.775  & $^{58}$Co  & 0.990     & 0.820 \\
834.838  & $^{54}$Mn  & 0.999 & 0.831 \\
911.196  & $^{228}$Ac & 0.262    & 0.656 \\
968.96   & $^{228}$Ac & 0.159    & 0.660 \\
1001.03  & $^{234m}$Pa & 0.0085 & 0.665 \\
1120.29  & $^{214}$Bi & 0.149   & 0.634 \\
1173.23  & $^{60}$Co  & 0.999   & 0.666 \\
1332.49  & $^{60}$Co  & 0.999   & 0.621 \\
1460.88  & $^{40}$K   & 0.106   & 0.626 \\
1764.49  & $^{214}$Bi & 0.153   & 0.533 \\
2614.51  & $^{208}$Tl & 0.358   & 0.372 \\
\noalign{\smallskip}\hline
\end{tabular}
\end{table}

\begin{figure}[h]
\centering
\includegraphics[width=0.97\columnwidth]{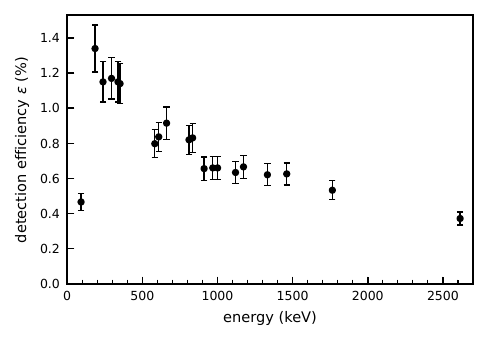}
\caption{Simulated detection efficiencies of Gator for the 16  R17317 PMTs as a function of the selected $\gamma$-ray energies. Error bars show the 10\% systematic uncertainty.}
\label{fig:efficiency_vs_energy}
\end{figure}

The sample's activity is determined from the full-energy deposition peaks in the spectrum. The detection efficiency depends on various factors such as the sample geometry, gamma-ray attenuation within the sample and support structure as well as the crystal size and thickness of its dead layer. This is evaluated using Monte Carlo simulations as detailed in section\,\ref{sec:efficiency}. 
Table\,\ref{tab:pmt_efficiencies} lists the obtained detection efficiencies and used branching ratios of the 16 PMTs with low radioactivity stems. The energy dependence of the efficiency is shown graphically in Figure\,\ref{fig:efficiency_vs_energy}.

\vspace{-\baselineskip}

\end{appendices}

\end{document}